\documentstyle[11pt,aaspp4]{article}

\def\spose#1{\hbox to 0pt{#1\hss}}
\def\lta{\mathrel{\spose{\lower 3pt\hbox{$\mathchar"218$}}
     \raise 2.0pt\hbox{$\mathchar"13C$}}}
\def\gta{\mathrel{\spose{\lower 3pt\hbox{$\mathchar"218$}}
     \raise 2.0pt\hbox{$\mathchar"13E$}}}

\begin{document}
\title{M84 - Can the AGN Affect The Orientation of The Disk?\altaffilmark{1}}

\author{
A.~C.\ Quillen\altaffilmark{2} \&
Gary A.\ Bower\altaffilmark{3} 
}
\altaffiltext{1}{Based on observations with the NASA/ESA {\it Hubble Space Telescope}
obtained at the Space Telescope Science Institute which is operated by the 
Association of University for Research in Astronomy, Inc. (AURA), under 
NASA contract NAS5-26555.}
\altaffiltext{2}{University of Arizona, Steward Observatory, Tucson, AZ 85721; 
aquillen@as.arizona.edu}
\altaffiltext{3}{Kitt Peak National Observatory, National Optical
Astronomy Observatories, P.O. Box 26732, Tucson, AZ 85726; gbower@noao.edu.}

\begin{abstract}
In M84 dust features are not aligned with the galaxy isophotes but are 
perpendicular to the radio jets.    We estimate the timescale
that the gas disk should precess in the galaxy.
Since this timescale is short ($\sim 2\times 10^7$ years at 100 pc)
we consider mechanisms that could cause the gas disk to be misaligned
with the galaxy.  One possibility is that a cooling flow 
replenishes the disk on this timescale and at angles that vary with radius.
Another possibility is that there is a pressure on the disk
that is large enough to overcome the torque from the galaxy.
We estimate this pressure and find that it can be provided by
ram pressure from an energetic interstellar medium that is consistent 
with velocity dispersions observed in ionized gas and densities estimated from
the X-ray emission.  We therefore propose that an AGN 
associated energetic interstellar medium is responsible for 
causing the gas disk in M84 to be misaligned with the galaxy isophotal 
major axis for $r \lta  600$ pc.   
We propose that AGN associated outflows or kinetic motion in low density media
could be responsible for jet/disk alignments observed on 100 pc scales
in nearby radio galaxies.

\end{abstract}


\section {Introduction}
Recent HST imaging and spectroscopic studies of active elliptical galaxies
have established that there are gas disks in the central
few hundred pc of these galaxies which could be feeding 
the massive $\sim 10^9 M_\odot$ black holes  at their centers
(e.g M84; \cite{bow97a}, \cite{bow97b}, M87; \cite{har94} and NGC 4261; 
\cite{fer96}).  
These gas disks are observed to be nearly perpendicular to the jets, 
establishing a link between the angular momentum of the disk and the 
direction of the jet.  
\cite{kot79} performed a survey of the observed angular 
difference, $\Psi$, between radio and disk axes 
in radio ellipticals and found a statistically significant 
peak in the distribution at $\Psi = 90^\circ$.  This established 
that radio jets are generally perpendicular to gas disks on large
(kpc) scales in nearby radio galaxies, even though 
no correlation between the radio axis and the galaxy isophotal major
axis exists at low redshift (\cite{san87}).
The strong peak at $\Psi = 90^\circ$ has been confirmed at 
smaller (100 pc) scales by \cite{van95_prime} using HST images.

Strangely, the jet direction is not necessarily expected to be correlated 
with the angular momentum of the gas feeding the black hole 
(and so the orientation of the disk)
because of Lense--Thirring precession, otherwise known
as `dragging of inertial frames' (\cite{ree78}).
A spinning black hole causes a gas disk 
near the gravitational radius, $r_g = GM/c^2 \sim 10^{14}$ cm, 
to precess, and so the 
disk or torus near the black hole is expected to fill a volume
that is axisymmetric and aligned with the spin axis of the black
hole itself (\cite{ree84}, \cite{bar75}).  
This inner torus is the proposed site for jet collimation and acceleration
(\cite{ree82})
so we expect that the jet should be aligned with the spin axis 
of the black hole but
not necessarily the angular momentum of the disk well outside of $r_g$.
A massive spinning black hole can be described as an angular momentum 
reservoir that does vary in momentum but only very slowly 
with the influx of fuel.
The rate of change of angular momentum is expected to be particularly
slow ($\sim 10^8$ years) in radio galaxies where the black holes 
are massive ($\sim 10^9 M_\odot$)
and the fueling rates are probably low ($\sim 10^{-4} M_\odot$/yr;
\cite{ree78}).  
This reservoir could also be responsible for stabilizing jets (\cite{ree78}).
If the black hole does not spin significantly, then
the jet axis would be determined
by the angular momentum of the inner disk. This disk, because of its
lower mass, could vary in orientation on shorter timescales than
the black hole axis (for a spinning system).
However, the observed alignment of jet and disk axes would
then require that the orientation of this inner
disk be coupled to that of the outer disk (at $\gtrsim 100$pc).  
Possible mechanisms for the alignment of the jet with the disk 
at 100 pc scales (well beyond $r_g$) have not been explored.

To explore alignment mechanisms, 
we consider the geometry of the warped dust features and ionized 
emission in M84 (NGC 4374, 3C 272.1), a radio elliptical in the Virgo cluster.  
M84 has a dusty disk seen in extinction in optical images
(see Fig. 1)  and in emission
in H$\alpha$ +[NII] that although nearly perpendicular to the jet,
is misaligned with the galaxy isophotes
(\cite{han85}, \cite{van95}, \cite{jaf94}, \cite{bow97a}, \cite{bau88}).
\cite{van95_prime} noted that the timescale for the disk to settle
into the galaxy plane of symmetry is probably short; $\sim 10^8$ years.
These authors therefore suggested that the misalignment could be explained by 
a rapid inflow of gas into the central region of the galaxy
(to radii smaller than $r \lta 600$ pc).
However if this is the case, it is difficult to explain the 
S-shape of the H$\alpha$ + [NII] emission (\cite{bau88}).
This emission is elongated and lies within $20^\circ$ of PA $\sim 70^\circ$ 
for $r \lta 7''$ but at this radius the disk twists to
PA $\sim 115^\circ$ on either side of the nucleus.
It is difficult to imagine how a cooling flow could generate 
such a morphology.
As an alternative to this possibility, we also consider an interaction between
the gas disk and an energetic interstellar medium (ISM) 
as a possible mechanism for aligning 
the disk perpendicular to the jet.

Throughout this paper we adopt a distance to M84 of 17Mpc
(\cite{mou95}; $H_0 = 75$km s$^{-1}$ Mpc$^{-1}$).
At this distance $1''$ corresponds to 82 pc.

\section{Misalignment of the Dust Features with the Galaxy}
That the dust features are misaligned with the galaxy isophotes
in M84 for $r \lta 600$ pc was emphasized by \cite{van95_prime}.
The galaxy isophotes have a major axis PA $= 129^\circ$ (\cite{van95};
see Fig. 1) whereas the dust features are elongated along PA$=80-65^\circ$  
(for $r<7''$) depending upon the region used to estimate the angle
(\cite{bow97a}, \cite{van95}).  
This is a misalignment of $\Psi = 50-65^\circ$
of the dust features with respect to the galaxy major axis.

Misalignment between gas and stellar isophotes  
is not necessarily rare, (e.g. see \cite{van95}). 
When there are kpc scale gas/galaxy misalignments usually 
the misalignment is the result of a merger 
(e.g. \cite{tub80} and \cite{ste92_3}).  
A gas disk in a non-spherical but axisymmetric galaxy will 
precess about the galaxy axis of symmetry.
At a radius of a few kpc the time for a gas disk to precess
about the galaxy axis of symmetry is long, $\sim 10^8$ years, requiring
a merger event to have occurred within this time if a merger is
responsible for the gas/galaxy misalignment.
Because the precession is faster in the center of the galaxy
in the central regions the warp can be multiply
folded and in the outer regions  where the precession is slower
the angle of the disk 
is more directly related to the orbital angular momentum of the merger event
(\cite{qui93}).  

After some time an initial planar disk that is precessing
freely will become multiply warped and the dust in the inner
region will form a
band of absorption that is roughly perpendicular to the galaxy axis of symmetry
(e.g. \cite{ste92}, \cite{spa96}, \cite{qui93_prime}, \cite{tub80}).
This is expected since a gas ring in one precession period
will trace out a cylindrical surface 
that is axisymmetric and perpendicular to the galaxy axis of symmetry.
The morphology of the dust lanes in M84 resembles those
of Cen A and NGC~4753 which are caused by 
multiply folded thin dusty surfaces (\cite{qui93_prime}, \cite{ste92}).
However precession in the galaxy potential is an unlikely explanation
for the warp in M84 since the band of absorption in the multiply
folded region is not aligned 
with the galaxy isophotes as is true in Cen A (\cite{tub80}) 
and NGC~4753 (\cite{ste92}).

In M84 the position angle of the dust features and ionized emission
varies with radius.  At $r\sim 4''$ the dust features are nearly
perpendicular to the radio jet, but at $r\sim 1''$ there is a line
of extinction north of the nucleus at PA $\sim 65^\circ$ (see Fig. 1).
The H$\alpha$ emission has PA $=58^\circ$ for radii $0''.25 -0.''5$
(\cite{bow97a}) however a disk at this angle did not yield a good
fit to the velocity profiles.  A good fit was achieved with
a disk more nearly perpendicular to the jet at PA $\sim 80^\circ$
(\cite{bow97b}).
The jet has PA $\approx 10^\circ$ on the pc scale and
PA $\approx 0^\circ$ on the 100 pc scale (\cite{jon81_3}).

\section{The Precession Timescale}

\cite{van95_prime} emphasized that the timescale for the gas disks to settle 
into the plane of the galaxy is short in the central few hundred pc.  
The timescale for a gas ring to precess about the galaxy axis
of symmetry, is shorter still.
The angular precession frequency $d\alpha/dt$ about the galaxy axis of 
symmetry for a gas ring of radius $r$ 
and inclination $i$ with respect to the galaxy axis of symmetry is 
\begin{equation}
{d\alpha \over dt} = \epsilon_\Phi \Omega \cos{i}
\end{equation}
(e.g. \cite{gun79})
where  $\epsilon_\Phi$ is the ellipticity of the gravitational potential 
which we assume is axisymmetric and
$\Omega \equiv v_c/r$ for $v_c$ the velocity of a particle 
which has only a gravitational force on it in a circular orbit.
Except in the case of a polar ring ($i \sim 90^\circ$),
the dependence on $i$ is weak.  
In a time 
\begin{equation}
t_p \sim {3 \over \epsilon_\Phi \Omega}
\end{equation}
which we refer to as a precession time,
rings that differ in radius
by a factor of 2 should have an angular difference of $\sim \pi/2$.
Here we have omitted the dependence on $i$.
After this time, an initially planar disk which is observed to be
aligned with the galaxy major axis at $r$, 
will at $r/2$ be maximally misaligned with this axis 
(e.g eqn. 7.4 of \cite{qui92}).
Since in M84 the dust is misaligned
with the galaxy for $r \lta 600$ pc, a mechanism operating at
a timescale faster than the above timescale, $t_p$, must be causing the
disk to remain at an angle differing from the galaxy isophotes.

To calculate the precession timescale we need to estimate 
the circular velocity, $v_c$.  This velocity can be estimated 
using Eqn. (4-55) of \cite{B+T_prime} from the 
observed stellar velocity dispersion ($\sigma_* = 303 \pm 5$km/s 
for $r\lta 5''$ \cite{dav88}) 
assuming that the galaxy is axisymmetric.
For a mild anisotropy of $\beta=0.4$ 
(where $\beta \equiv 1 - {v_\theta^2 \over v_r^2} $ is the ellipticity
of the velocity ellipsoids), the circular velocity would be 
$v_c \sim 1.1 \sigma_* \sim 330$km/s (where we have assumed that
$\rho \propto r^{-2}$ consistent with the small variation in $\sigma_*$
with radius).  A smaller anisotropy would result in a higher circular velocity.
The minimal galaxy rotation ($<8$km/s; \cite{dav88}) 
suggests that the axis ratio and anisotropy are not high 
(using the virial equations).

To calculate the precession rate we must also estimate the
ellipticity of the gravitational potential, $\epsilon_\Phi$,
which is directly related to the ellipticity of the isophotes (see Fig. 1).
Here the ellipticity $\epsilon \equiv 1- b/a$ where $b/a$ 
is the axis ratio of the isophotes. 
In M84 \cite{van95_prime} measured an ellipticity of $0.21$ between
$10''$ and $12''$ and \cite{fer94_prime}
measured $0.187\pm 0.002$ for $5<r<15''$.
Both of these works found 
slight decreases in the ellipticity with increasing radius at $r<15''$, 
which extends into the outer parts of the galaxy ($r>100''$) 
where $\epsilon < 0.1$ (\cite{pel90}). 
Only minimal deviations from the shape of a pure ellipse were 
observed (\cite{vdb94}, \cite{pel90}).
Measurement of the isophote ellipticity in the optical images 
is complicated by the presence 
of the dust features in the inner region (see Fig. 1).
In the near-infrared 2.2$\mu$m (Ks) image of M84 from \cite{pah94} 
the isophotes are completely symmetrical, do not vary in position angle, 
and show little extinction from dust.
The ellipticity from this image agrees with the other measurements and
show that there is no change in ellipticity to $r \sim 2''$, but it
does not have sufficient spatial resolution (seeing $\sim 1''$)  
to accurately measure it for $r\lta 2''$.
A higher resolution near infrared image may be able to see if the ellipticity
extends to smaller radii.

From this we now estimate the ellipticity of the gravitational potential.
For power-law potentials (with a core)
$\epsilon_\Phi \sim \epsilon_\rho/3$ (see \cite{B+T}, Figure 2-13),
where $\epsilon_\rho$ is the ellipticity of the density distribution.
Using $\epsilon_\rho = 0.15$ close to that measured from the isophotes
we estimate $\epsilon_\Phi \sim 0.05$.   This is actually a lower limit
since projection causes the measured ellipticity 
to be less than $\epsilon_\rho$ and in the core $\epsilon_\Phi$ 
could be higher than $\epsilon_\rho/3$.

With the above values for $\epsilon_\Phi$ and $v_c$ we estimate a 
precession time of
\begin{equation}
t_p = 2 \times 10^7~{\rm yr}
\left({0.05 \over \epsilon_\Phi}\right) 
\left({r \over 100 {\rm pc}}\right) 
\left({330 {\rm km/s }\over v_c} \right).
\end{equation}
Since in M84 the dust is misaligned
with the galaxy for $r \lta 600$ pc, a mechanism operating at
a timescale faster than the above timescale must be causing the
disk to remain at an angle differing from the galaxy isophotes.

In the above discussion we have assumed that M84 is axisymmetric, 
and not triaxial.
The cuspiness observed in the isophotes of many ellipticals (\cite{geb96}),
including M84 (\cite{fer94}), coupled with theoretical work measuring
mixing timescales for stochastic orbits in cuspy potentials (\cite{mer96})
suggest that triaxiality can only be short lived in the central regions
of M84.    
There is little twist in the isophotes of M84 (\cite{vdb94}, \cite{pel90})
which should be observed if the galaxy is strongly triaxial and not oriented
with an axis coincident with the line of sight.  
In addition \cite{van95_prime} found
that triaxiality could not explain the observed disk/galaxy misalignments 
in a sample of elliptical galaxies.  Whereas precession times
are similar, disk settling times in triaxial
galaxies are substantially faster than in axisymmetric systems 
(\cite{hab85}).
The possible warped equilibrium state
in a tumbling triaxial galaxy (\cite{van82}) is unlikely
because there is little stellar rotation observed in M84 (\cite{dav88}).

\section{What Operates Faster Than Precession?}
Here we consider various mechanisms operating on timescales
faster than the precession timescale that can cause the gas disk to
be misaligned with the galaxy.


The mass deposition rate from the cooling flow estimated by \cite{tho86} 
for $r<1$ kpc is $0.2 M_\odot$/yr. 
At this rate the gas disk with mass $\sim 10^6 - 10^7 M_\odot$
(\cite{van95}, \cite{bow97a}) could be replenished in $\sim 10^7$ years. 
This is sufficiently fast that if the cooling flow has some angular momentum
the angle of the disk could be different than that of
the galaxy.  However, mass lost
from stars in the galaxy should not be rotating since the stars 
have no observed
net rotation (\cite{dav88}).  The cooling flow could perhaps
be influenced by the cluster environment, and thus gain some angular momentum.
If the cooling flow quickly replenishes the gas disk then we would 
expect that it 
would have orientation that does not vary with radius 
and so should not exhibit the S-shape noted by \cite{bau88_prime} 
in H$\alpha$ + [NII] emission (see Fig. 1).

If the galaxy disk has a large radial inflow rate then a quasi-stationary
warp could develop (e.g. \cite{ste88}) where the gas flows in faster
than it can turn by precession.  A radial inflow velocity greater than 
$v_r >  v_c  \epsilon_\Phi  \approx 16 $ km/s is needed for this to occur.
This inflow rate is so fast that the mass inflow rate would be much higher than
that needed to power the AGN, requiring some place
to put the accreted mass, such as a wind, a dense inner disk, 
or advection-dominated accretion into the black hole (e.g.~\cite{abr88}, \cite{abr96}).  
Also the lifetime of the disk 
would then be of the same time as the precession timescale $t_p$ 
either requiring a substantial gas reservoir at radii larger than 600 pc or
implying that the lifetime of the disk is only a few times $10^7$ years.
Such high inflow rates could only be achieved in an isolated accretion
disk with a very high viscosity and velocity dispersion.
However, if the cooling flow were contributing low angular momentum
gas to the disk (e.g. \cite{gun79}) a thin lower dispersion disk might be able
to sustain such high inflow rates.    
In this case some precession in the galaxy potential could occur.  
It is unclear if this scenario could account for the observed 
disk morphology.

Recently \cite{pri96} has proposed that radiation from an AGN can drive 
a warp in an accretion disk.  However this mechanism operates only when 
the force from 
radiation pressure is significant compared to the gravitational force.
\cite{mal96} find that only in the Keplerian regime in
particularly luminous AGN can this occur.
We roughly estimate that the AGN in M84 would
have to be $100$ times more luminous 
(the radio bolometric luminosity is $6\times 10^6 L_\odot$; \cite{her92})
for a radiatively driven warp to operate on an optically
thick disk at 100 pc.

\subsection{Pressure Required to Overcome the Galaxy Torque}

Here we consider the possibility that 
the misalignment of the disk with the galaxy is caused locally.
There is ample observational evidence for AGN associated outflows 
and energetic ISM both at high and low redshift.  
$10-20\%$ of optically selected quasars, known as broad absorption line
quasars (BALs), show blueshifted broad absorption lines (\cite{wey91})
in the same resonance lines that are seen in emission in most QSO spectra
(including the BALs).  In nearby active systems there is evidence
for complex velocity structure and velocity gradients suggestive of outflow
across the narrow line region (NLR) (NGC 1068; \cite{eva94}, 
NGC 5252; \cite{aco96}, Markarian 1066; \cite{bow95}).

AGNs are expected to impart kinetic energy to the surrounding media.
\cite{mur95} propose that a wind ``can be driven up out of
the disk by a combination of radiation pressure and gas pressure''.
Whittle (1992a) and (1992b) finds that [OIII] lines are significantly
broader in Seyferts with a high radio luminosity, suggesting
that jets also accelerate clouds in the narrow line region.
HST images of the narrow line region in some Seyfert galaxies show
conical features in emission lines interpreted to be ionized
gas entrained in a jet,
or material ionized by an anisotropic or shielded central source 
(e.g. \cite{sto92}).  
Jets can cause large scale shocks in the ISM of a galaxy 
(\cite{dop95}, \cite{ped85}), 
as has been observed in NGC~4258 (\cite{cec95}) and Mkn 573 (\cite{pog95}).

Radio galaxies put a substantial fraction of their luminosity
into the kinetic energy of their jets (e.g. \cite{ree82}).
Emission in H$\alpha$ + [NII] commonly shows 
high velocity dispersions of the order of a few hundred km/s 
(\cite{bau92_3}, \cite{tad89_3}, \cite{axo89}).
In radio galaxies there is strong evidence that the jets themselves 
impart significant motions in the surrounding ISM.  
In Cygnus A and 3C~265, two narrow line emitting 
components near the jets are separated by velocities as large as
1600-1800 km/s (\cite{tad91}).  
Large line widths have also been observed in Seyferts with jets (\cite{cap95a}, 
\cite{axo97}, \cite{cap95b}). 
If the jets are responsible for motions in the surrounding
ISM then there should be a differential in this medium, with
the largest motions nearest the jets, and the lowest 
motions in the plane perpendicular to the jets.

If the disk orientation is affected by the local medium then 
there must be a pressure on the gas disk greater than 
that of the gravitational potential which would be causing it to precess.
The torque per unit mass on a ring of radius $r$ is
\begin{equation}
\tau =  \epsilon_\Phi \cos{(i)} v_c^2 
\end{equation}
and so the pressure required to keep the disk from precessing is 
\begin{equation}
P > \epsilon_\Phi \cos{(i)} \Sigma v_c^2 /r 
\end{equation}
where $\Sigma$ is the mass per unit area in the gas disk.
If we assume that a hot gas with density
$\rho_h$ and motions with velocity $v_h$ exerts a 
ram pressure $\rho_h v_h^2$ on the disk then given the density of the hot gas
surrounding the disk we can estimate the velocity dispersion, $v_h$, 
required to overcome the galaxy torque.
From the X-ray emission, \cite{tho86} find an electron density that is 
$n_e \approx 0.5  (100{\rm pc}/r) {\rm cm}^{-3}$ 
(where we have extrapolated from their profile which ranges 
from 0.5 -- 20 kpc).
From this we estimate 
\begin{equation}
v_h \approx 
50 {\rm km/s}
\left({\epsilon_\Phi \over 0.05}                     \right)^{1/2}
\left({\Sigma        \over 1 M_\odot/{\rm pc}^2}     \right)^{1/2}
\left({0.5 {\rm cm}^{-3}    \over n_e(r=100{\rm pc})}      \right)^{1/2}
\left({v_c           \over 330 {\rm km/s}       }    \right)
\end{equation}
where we have assumed that $n_e \propto 1/r$ and ignored the weak
dependence on inclination.

The random velocity component required depends on the surface density,
$\Sigma$, 
of the disk which is difficult to estimate.  From the extinction in the
dust features, \cite{van95_prime} and \cite{bow97a_prime} estimate a total
disk mass of $10^6 M_\odot$  and $9 \times 10^6 M_\odot$ respectively.
For a disk of constant surface density truncated at $r=400$ pc these 
mass estimates
give $\Sigma = 2 - 18 M_\odot {\rm pc}^{-2}$.  We expect that the density
would increase with decreasing radius, and so that the density
would be higher than these values at small radii and lower at large radii.
The resulting velocity $v_h$ required to overcome the torque from the galaxy 
could be a few times higher than the above 50 km/s.

Is there evidence for an energetic medium in M84 that has
velocities of a few hundred km/s?
The profiles in [NII] by (\cite{bow97b})
have a FWHM $\sim 200$km/s, so that 
the velocity dispersion observed in the ionized gas is quite high,
even when observed at their high angular resolution ($0.1''$). 
This suggests that the high dispersion is not an artifact caused by
`beam smearing' or by sampling a large region in a cold slowly rotating disk.
In the lower angular resolution ($\sim 1''$) observations of 
\cite{bau90} the velocity dispersions are $\sim 150$km/s  along
the major axis of the H$\alpha$+[NII] emission increasing to 300km/s 
in the central few arcsecs.  
Dispersions measured along the minor axis are higher still
with values of 200--400 km/s.  We note that along the minor axis 
in a rotating disk, artifacts
caused by beam smearing should be minimal because
the line of sight velocity is equal to the systemic velocity of the system.
The higher dispersion along the minor axis 
(which also corresponds to the radio jet axis) suggests that motions
are largest nearest the jets.   

As described by \cite{bow97a_prime} 
the H$\alpha$ + [NII] appears to have 3 components:
a rotating disk, an ionization cone along the radio jet axis, and outer 
filaments that coincide with emission seen in the image of \cite{bau88_prime}.
The [NII] velocity profiles of \cite{bow97b_prime} and
\cite{bau90} suggest that these components contain complex velocity structure
with motions of few hundred km/s.
We conclude that there is evidence for an energetic ISM with motions and density
large enough that this medium could contribute sufficient ram pressure 
to overcome the galactic torque.

Even though \cite{bow97b_prime} and \cite{bau90} observe velocities consistent
with rotation in a disk along the major axis
in the H$\alpha$ + [NII] emission, the rotational velocity ($<200$km/s)
is substantially lower than $v_c \sim 330$km/s which we estimated above.
The axis ratio of the emission is consistent with a
nearly edge-on disk which therefore would be observed at the full
rotational velocity if undergoing circular motion.
Because the observed rotational velocity is low
there must be non-gravitational forces on the ionized gas.
For a disk containing clouds with random motions the mean rotational
velocity is expected to be lower than $v_c$ 
because of the radial pressure force on the disk, (similar in nature
to asymmetric drift, e.g. see \cite{B+T}, Eqn. 6-24).
However if this force is large enough to significantly lower 
the rotational speed then radial inflow would necessarily be rapid,
and the disk should be quite thick.  Hydrostatic equilibrium in a nearly
round gravitational potential requires that 
${h \over r} \sim {\sigma \over v_c}$
for $h$ the thickness of the disk and $\sigma$ the velocity dispersion
of clouds in the disk.
The velocity dispersion of the H$\alpha$ emission contrasts sharply 
with the sharpness of the dust features seen in absorption
(Fig. 1). This sharpness suggests that the disk
is quite thin and so implies that the velocity dispersion in the cold 
gas associated with the dust cannot be large
($\lesssim 30$ km/s for $h/r \lesssim 10$).    
This is strong evidence that we are not observing
the same material in H$\alpha$+[NII] as in absorption from dust.
If the pressure gradient in the ionized
disk itself is not sufficient to cause the low disk rotational velocity, 
then it is more likely that the low rotation seen in the 
H$\alpha$ + [NII] disk is a symptom
of an interaction between a non-rotating energetic medium and the rotating
cold disk.  
\cite{gun79_prime} considered the interaction of a cold gas disk in an elliptical
galaxy containing a hot X-ray emitting gas.
He pointed out that since the X-ray gas is pressure supported, it 
should not be rotating.  This causes a velocity 
sheer between the hot gas and the cold disk which would result 
in Kelvin-Helmholtz
instabilities, and an additional force on the disk which could lower its 
rotational velocity and increase its accretion rate.


\section{Summary and Discussion}

In this paper we have estimated the timescale of a gas disk 
to precess in the non-spherical gravitational potential of M84.
This timescale is a few times $10^7$ years at 100 pc where the dust features
are misaligned with the galaxy isophotes.
For the disk to remain misaligned with the galaxy potential some mechanism
must operate faster than this.
While a cooling flow could replenish the disk on this timescale
it is difficult to explain why the disk is at a roughly constant
angle within $r<7''$ and yet twists at this radius forming an overall S-shape 
in the H$\alpha$ + [NII] emission
(\cite{bau88}).  Extremely fast accretion through the disk itself 
would require a substantial gas reservoir at large radii or a very
short disk lifetime. 
It would also require a place to put the excess accreted
gas mass, such as a wind, an inner disk or advection-dominated accretion
into the black hole.    
A combination of fast accretion and replenishment
by a cooling flow could possibly result in 
inflow faster than the precession rate, but it is not clear whether this
combination could account for the disk morphology.
The AGN is not luminous enough for 
the radiative induced warp mechanism of \cite{pri96} to operate.

As an alternative to these external mechanisms we 
consider the possibility of a local (presumably AGN associated)
force on the disk.  We estimate the ram pressure required to overcome the torque
from the galaxy and find that energetic motions of the scale observed
in the ionized gas and the density inferred from
the X-ray gas could provide this pressure.
That the velocity dispersion seen in H$\alpha$ + [NII] emission
is higher along the jet than along the disk major axis suggests that there is 
a pressure differential in the hot medium 
with motions lowest in the plane perpendicular to the jet.  
The low rotational velocity of the H$\alpha$ + [NII] emission implies 
that there are significant non-gravitational forces on the gas.
We therefore propose that an energetic low density medium in M84
is responsible for causing the gas disk to be perpendicular to the jet
on the scale of a few hundred pc.

The pressure required to overcome the galactic torque 
depends on the ellipticity of the potential, and 
it is difficult to measure the ellipticity of the 
isophotes from optical images because of the dust features themselves.  
For example at small radii
where the density of the disk is expected to be highest 
the galaxy could be nearly spherical.  At $r\sim 1''$, the
galaxy surface brightness profile has a shoulder or cusp (\cite{fer94}).
If this shoulder has been caused by the black hole (\cite{you80}) then 
this radius would be a likely place for a change in the ellipticity
of the galaxy.  
A higher resolution near infrared image may be able to see if the ellipticity
extends to smaller radii.

We note that the X-ray observations of \cite{tho86} 
did not have sufficiently high angular resolution to see
if the electron density profile is a power law all the way
into the nucleus.  The cooling time 
($\propto n_e^{-1}$) of their last measured
point at $r=500$ pc is only $\sim 2 \times 10^7$ years which suggests 
that substantial mass and energy input is required near the nucleus
if the electron density increases all the way into the nucleus.
Higher resolution X-ray observations could determine if this is
the case.   Pressures estimated from the [SII] 6717, 6731\AA~ lines 
in cooling flow galaxies (including M87 within $r <200$ pc) 
in the central regions are high, $\sim 1-2 \times 10^{-9}$ dynes cm$^{-2}$ 
(\cite{hec89})
and not inconsistent with high central densities extrapolated from the
lower resolution X-ray estimated pressures.
Higher spectral resolution observation of these [SII] lines could
therefore be used to estimate the pressures at small
radii in the $10^4$K gas (which should
be nearly in near pressure equilibrium with the X-ray emitting $10^7$K gas).
If there is energetic dense gas mixed in with the X-ray
emitting gas, then this medium could exert a higher pressure 
on the gas disk than the hot gas alone.

The velocity profiles of a large region of the disk could 
determine the orientation of the disk as a function of radius, 
as well as the velocity
structure of the ionization cones and filaments.
The extent that motions in this medium are ordered rather than random 
could be studied with such a data set.  If the jet itself is a 
source of kinetic energy input into the low density medium, then 
the velocity structure of gas nearest the jet should differ from
that outside it.  
If there is indeed a large pressure from an energetic ISM on the gas
disk in M84 then we would expect shocks, associated heating and mixing
layers.  The location and conditions of these processes 
can be probed with emission line diagnostics.

Recent investigations find that almost all ellipticals have 
dust (\cite{vdb94}).  In cases of non-active elliptical galaxies 
having dusty disks misaligned with the galaxy isophotes, the misalignment
might be caused by another mechanism (perhaps a cooling flow or a
galaxy merger). van Dokkum \& Franx's (1995) sample contains only five
galaxies with radio power $\log P_{\nu}$ (6 cm)$ < 20$ (W/Hz)  and
misaligned dust features.  Further investigation is needed with a
larger sample to determine how common these non-AGN misaligned cases are.

If it is indeed true that AGN associated energetic ISM can change the 
orientation 
of a gas disk then it would be interesting to investigate correlations
between the AGN and the properties of the disk.
For example massive disks would require a larger force to align them.  
A more energetic and higher density ISM should be able to more easily align
a disk, suggesting that the radius past which a disk ceases to be
aligned with the jet should be correlated with the pressure
and velocities observed in the hot material and the radio
or jet luminosity.
The outer parts of the jet would be related to the
outer parts of the disk rather than the angular momentum of the outer
part of the disk determining the future alignment of the jet.
The outflow of an AGN could also affect its own accretion rate.
It could either destroy its own disk, or increase the velocity dispersion
and inflow rate in the accretion disk as has been proposed in the case of
the starburst/AGN interaction (\cite{von93}).

Hydrodynamic simulations of jets sometimes show backflows which result
in over-pressured jets but also cause higher pressures and 
energetic motions in the plane perpendicular
to the jet (e.g.~\cite{har90}).    These authors have
noted that the boundary condition in the plane perpendicular to
the jet can have a substantial effect on the jet hydrodynamics.  
One possibility is that the backflow 
is partly responsible for the ram pressure which we
propose here affects the alignment of the disk.
Jet simulations could be used to explore the forces on the disk
as a function of ISM and jet physical parameters.

\acknowledgments

R. Green provided the inspiration for this work as well as many
valuable discussions.
We also acknowledge helpful discussions and correspondence with 
G. Rieke, M. Rieke, G. Schmidt, D. Hines, P. Pinto, D. DeYoung and F. Melia. 
ACQ also acknowledges support from NSF grant AST-9529190 to M. and G. Rieke
and NASA project no. NAG-53359.  GB acknowledges support from the STIS 
Investigation Definition Team.

\clearpage


\begin{figure*}
\vspace{1.0cm}
\caption[junk]{
HST/WFPC2 continuum imaging of the center of M84 from 
Bower et al.~(1997a), with a resolution of 0.1" (8 pc). The
grayscale shows the (V-I) map, with values of (V-I) ranging linearly from 
2.0 (black) to 1.3 (white).  The reddest color in the data is (V-I) = 1.7. 
Contours from the V-band image are superimposed, where the contour 
interval corresponds to a factor of $\sqrt{2}$ in intensity.
The jet has PA $\approx 10^\circ$ on the pc scale and
PA $\approx 0^\circ$ on the 100 pc scale (Jones, Sramek, \& Terzian 1981).
The dotted lines show the approximate morphology of the S-shape twist
in the larger scale H$\alpha$ + [NII] emission map of Baum et al.~(1988).
Within these dotted lines the large scale emission is alligned with 
the dust lanes.
\label{fig:fig1} }
\end{figure*}

\end{document}